\documentclass{moriond}

\def\be{\begin{equation}}
\def\ee{\end{equation}}
\def\bea{\begin{eqnarray}}
\def\eea{\end{eqnarray}}

\usepackage{amsmath}
\usepackage{subfigure}
\usepackage{booktabs}

\newcommand{\Photo}{\includegraphics[height=35mm]{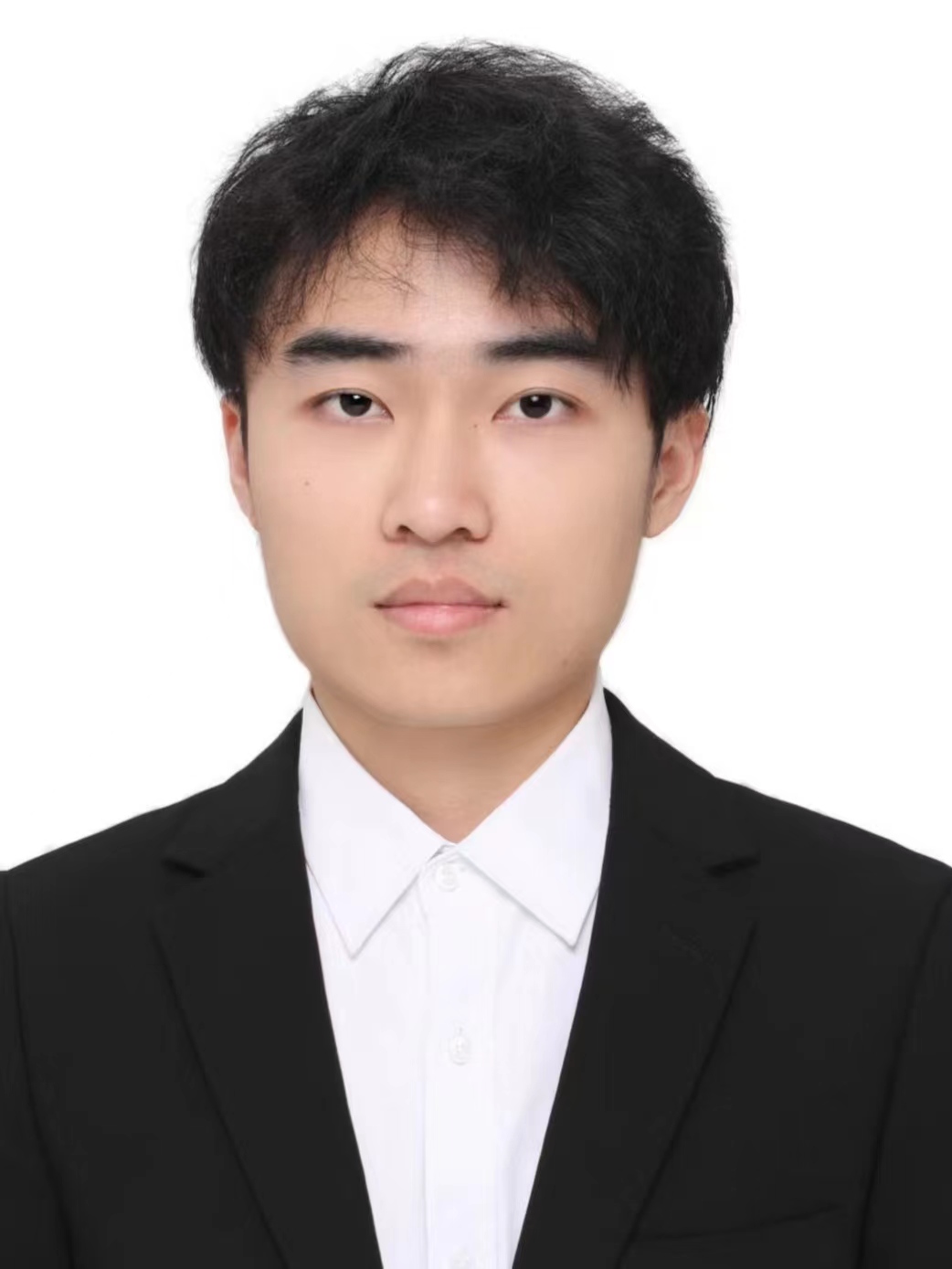}}

\begin{document}
\vspace*{4cm}
\title{Search for dark sector and rare decays at BESIII}

\author{ Zhi-Jun Li, Zheng-Yun You on behalf of BESIII }

\address{School of Physics, Sun Yat-sen University, Guangzhou 510275, China}

\maketitle\abstracts{
The BESIII experiment has collected a large data sample of charmonium, charm mesons, hyperons, and other light mesons. These data provide a unique opportunity to explore the dark sector and rare decays. We present recent dark sector results from the BESIII experiment, including searches for sub-GeV dark matter in $\eta \to \pi^0 + \text{invisible}$ and $J/\psi \to \phi + \text{invisible}$, searches for dark baryon particles in $\Xi^- \to \pi^- +\rm{invisible}$, and searches for light vector bosons in $\chi_{cJ} \to J/\psi X$, where $X \to e^+ e^-$. In addition, we also present the recent rare decay searches, including a search for $\psi \to e \mu$, as well as searches for several baryon number and lepton number violation processes, and searches for various charmonium weak decays.}

\section{Introduction}

The Standard Model (SM) has achieved remarkable success in explaining a wide range of phenomena in our universe, covering fundamental particles and their interactions. Nevertheless, numerous unanswered questions persist, including issues related to dark matter (DM), the strong CP problem, the muon $g-2$ anomaly, the hierarchy of fermion masses, and the asymmetry between matter and antimatter, among others. These unresolved issues indicate the potential existence of the dark sector beyond the SM, which may include new particles and interactions linking the dark sector with SM matter. The introduction of these new particles or interactions could also increase the rates of specific rare decays.
Exploring the dark sector and rare decays opens up an exciting pathway for investigating new physics (NP) beyond the SM. If dark sector particles have masses in the $\sim$GeV region, they could be probed through high-intensity $e^+e^-$ collider experiments, such as the Beijing Spectrometer III (BESIII)~\cite{BESIII:2009fln}.

BESIII is a general-purpose spectrometer developed for exploring $\tau$-charm physics at center-of-mass energies ranging from 1.84 to 4.95~GeV. It records symmetric electron-positron ($e^+e^-$) collisions produced by the Beijing Electron Positron Collider II (BEPCII) storage ring~\cite{Yu:2016cof}. To date, BESIII has collected extensive datasets within this energy range, which includes 10 billion $J/\psi$ events, 2.7 billion $\psi(2S)$ events, a 20~fb$^{-1}$ dataset collected at 3.773 GeV, and more than 20~fb$^{-1}$ of data obtained above 4.0 GeV. This substantial collection of charmonium data enables the production of large samples of hyperons and light mesons via charmonium decays. Leveraging these extensive datasets along with advanced analytical techniques~\cite{Li:2024pox}, BESIII is ideally equipped to conduct in-depth studies of the dark sector and rare decays.

\section{Search for sub-GeV DM in $\eta\to\pi^0+\rm{invisible}$}
DM is strongly motivated by astronomical observations; however, the SM of particle physics cannot explain its existence. Conventional approaches for the direct detection of DM often rely on nuclear recoils to search for elastic scattering events between DM particles and target nuclei. Strong constraints have been established for GeV-scale DM, while the detection of sub-GeV DM remains significantly more limited due to insufficient recoil energy to exceed the detection threshold. 
BESIII operates in the $\sim$GeV region and presents a unique opportunity to probe sub-GeV DM from the collider.

Considering that a dark boson mediates the interaction between DM and SM particles, the DM-nucleon scattering ($\chi N \to \chi N$) can be associated with the same NP as processes like meson decay (e.g., $\eta \to \pi^0 \chi \chi$), where $\chi$ represents the DM particle and $N$ denotes the nucleon. This commonality in physics enables a direct comparison between results from collider experiments and direct detection experiments. Among the various meson decay processes to DM, $\eta \to \pi^0 \chi \bar{\chi}$ is predicted to have the largest branching fraction (BF) for the same coupling strength, owing to its unflavored nature and narrow width. In the decay of $\eta \to \pi^0 \chi \bar{\chi}$, the dark boson is considered to be a scalar of $S$. Typically, new scalar bosons are assumed to exhibit Higgs-like couplings, which has resulted in relatively little experimental attention being directed towards searches for couplings to light states. However, some theoretical studies have indicated that a flavor-specific scalar boson can still possess a relatively sizable coupling to light quarks~\cite{Batell:2018fqo}, thereby emphasizing the potential for new physics in light meson decays.

At BESIII, $\eta$ mesons can be obtained from $(10\,084 \pm 44) \times 10^6~J/\psi$ decays, specifically in the channel $J/\psi \to K^+K^-\eta$, where the charged kaons assist in tagging the $\eta$ and reconstructing DM. The $\eta$ candidates are identified by selecting events with $K^+K^-$ and $\eta \to \text{anything}$. A total of $(2185.3 \pm 3.4) \times 10^3$ inclusive decay $\eta$ events are reconstructed. The search for invisible DM is further searched by requiring an additional $\pi^0$ in the event, which is reconstructed by the recoiling of $K^+K^-\pi^0$. This analysis specifically searches for on-shell decays $S \to \chi \bar{\chi}$.
No significant signal is observed in the mass range of $m_S = 0 \sim 400\,\text{MeV}$. The corresponding upper limits (ULs) on the BFs of $\eta \to \pi^0 S \to \pi^0 \chi \bar{\chi}$ are determined to be $(1.8 \sim 5.5) \times 10^{-5}$ at the 90\% confidence level (CL)~\cite{BESIII:2026qsu}.

\vspace{-0.0cm}
\begin{figure*}[htbp] \centering
	\setlength{\abovecaptionskip}{-1pt}
	\setlength{\belowcaptionskip}{10pt}

        \subfigure[]
        {\includegraphics[width=0.49\textwidth]{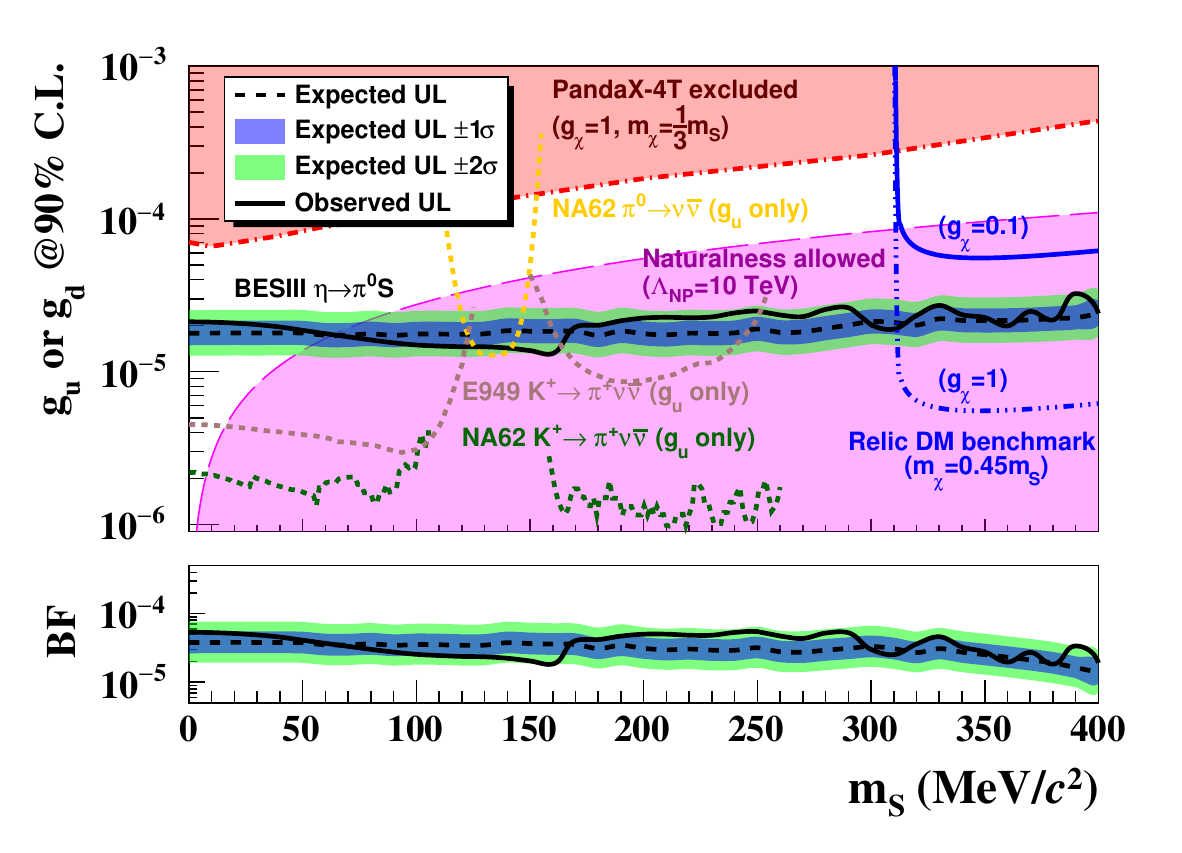}}
        \subfigure[]
        {\includegraphics[width=0.49\textwidth]{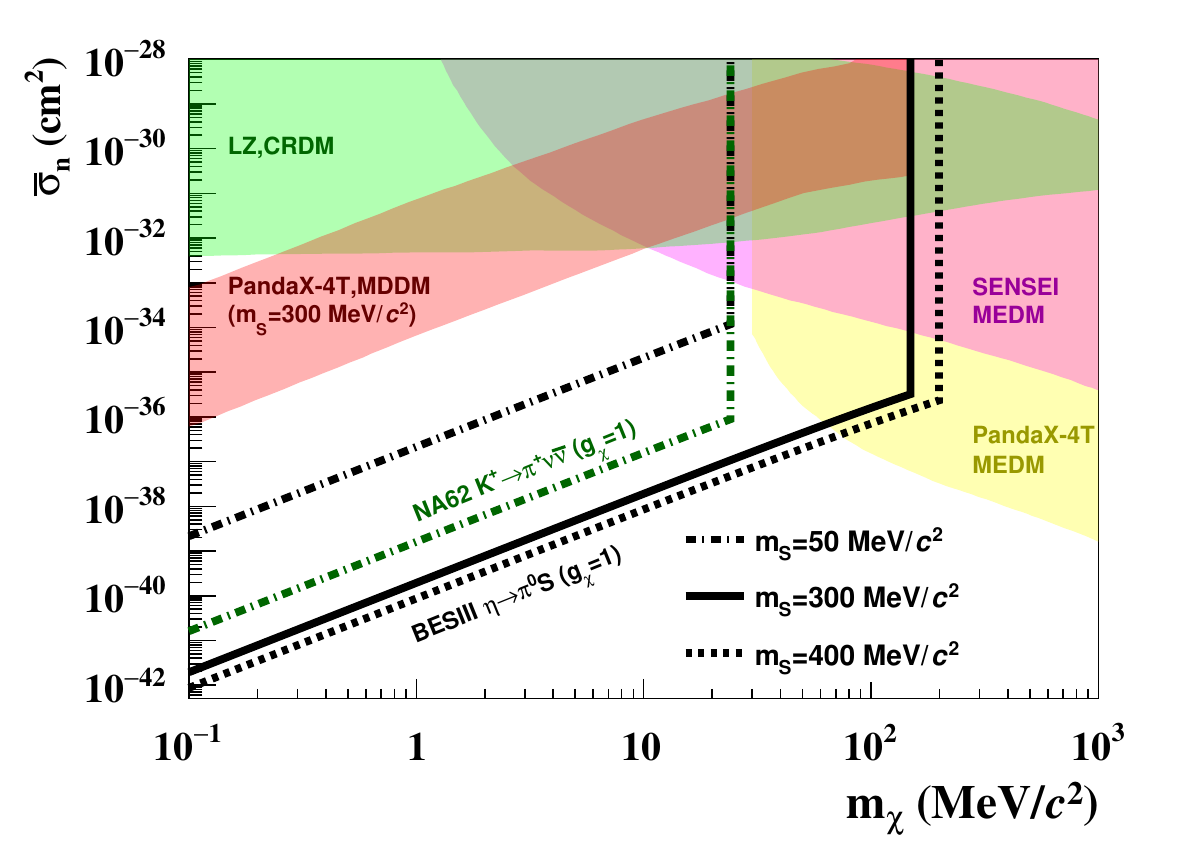}}
        
	\caption{
        (a) The ULs on the coupling strengths $g_u$ or $g_d$ (up) and the BFs of $\eta\to\pi^0S$ (bottom) for $S$ with different masses. The purple-filled region represents the naturalness allowed space in the EFT with $\Lambda_{\rm{NP}}=10~\rm{TeV}$ and the blue lines illustrate the thermal relic DM benchmarks for $g_{\chi} = 0.1$ and $g_{\chi} = 1$. (b) The constraints on $\bar{\sigma}_n$ from this work and direct detection experiments.
        } 
	\label{fig:eta2pi0}
\end{figure*}
\vspace{-0.0cm}

Considering the interaction Lagrangian of the flavor-specific scalar~\cite{Batell:2018fqo}:
\begin{eqnarray}
\mathcal{L} \supset -g_{\chi} S \bar{\chi}_L \chi_R - g_u S \bar{u}_L u_R + h.c.,
\end{eqnarray}
where $g_u$ ($g_\chi$) is the effective coupling strength between $S$ and the SM quark (DM), and $L(R)$ marks the chirality. The coupling to the down quark, with a coupling strength of $g_d$, is analogous. The constraints on $g_u$ or $g_d$ from the $\eta\to\pi^0S\to\pi^0\chi\bar{\chi}$ search are shown in Figure~\ref{fig:eta2pi0} (a), with the ULs of $(1.3\sim3.2)\times10^{-5}$ at the 90\% CL.
Compared to the previously published result from PandaX-4T~\cite{PandaX:2023tfq}, which measured the DM-nucleon scattering mediated by $S$, the constraint from BESIII is improved by a factor of 3.3 to 18.3. 
The constraints from $K^+\to\pi^++\rm{invisible}$ searches~\cite{NA62:2025upx,BNL-E949:2009dza,NA62:2020pwi} are also shown for comparison, but it is important to note that $K^+\to\pi^++\rm{invisible}$ is sensitive only to $g_u$, while $\eta\to\pi^0+\rm{invisible}$ is sensitive to both $g_u$ and $g_d$.

In Figure~\ref{fig:eta2pi0} (a), the naturalness allowed region is from the effective field theory~(EFT) with the relationship of $g_u \leq \frac{16\pi^2}{\sqrt{2}} \frac{m_S\nu}{\Lambda^2_{\rm{NP}}}$~\cite{Batell:2018fqo},
where $\Lambda_{\rm{NP}}$ is the NP energy scale, $\nu=246.2 \rm{GeV}$ is the Higgs vacuum expectation value.
The NP energy scale is set to be 10~TeV, illustrating that the sensitivity of the $\eta\to\pi^0S$ decay at low energy can probe the allowed space from a high NP energy scale. 
Considering freeze-out mechanism in the cosmic evolution, Figure~\ref{fig:eta2pi0} (a) also shows the thermal relic DM benchmark that accounts for the observed cosmological DM density of $\Omega_{\rm{DM}}h^2 = 0.12$ through the annihilation of $\chi\bar{\chi} \to S \to \pi\pi$, by assuming $m_{\chi} = 0.45 m_S$. For $g_{\chi} = 0.1$, $\eta\to\pi^0+\rm{invisible}$ search can probe and exclude the parameter space of thermal relic DM, whereas for $g_{\chi} = 1$, a larger dataset is still required.

Considering DM scattering with the nucleon by exchanging $S$ searched at BESIII, the constraints on the scattering cross section ($\bar{\sigma}_n$) are derived in Figure~\ref{fig:eta2pi0} (b), where $g_{\chi}$ is assumed to be 1, and a smaller $g_{\chi}$ would result in a more stringent constraint. The constraint on $\bar{\sigma}_n$ from BESIII is stronger by approximately 5 orders of magnitude over previous direct detection experiments, providing unique insights into sub-GeV DM.

\section{Other new dark sector searches at BESIII}

\subsection{Search for sub-GeV DM in $J/\psi\to\phi+\rm{invisible}$}
\begin{figure*}[htbp] \centering
	\setlength{\abovecaptionskip}{-1pt}
	\setlength{\belowcaptionskip}{10pt}

        \subfigure[]
        {\includegraphics[width=0.31\textwidth]{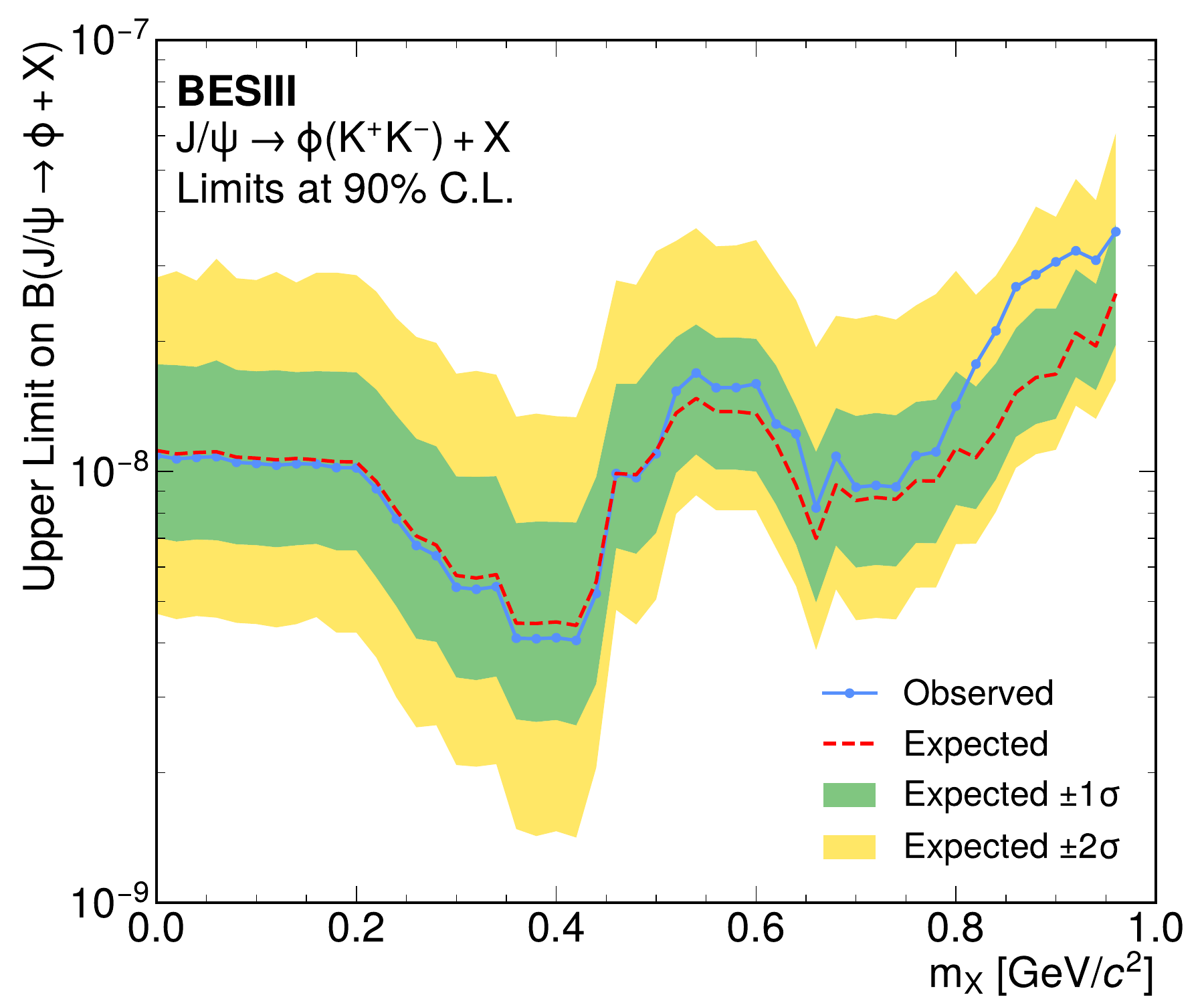}}
        \subfigure[]
        {\includegraphics[width=0.36\textwidth]{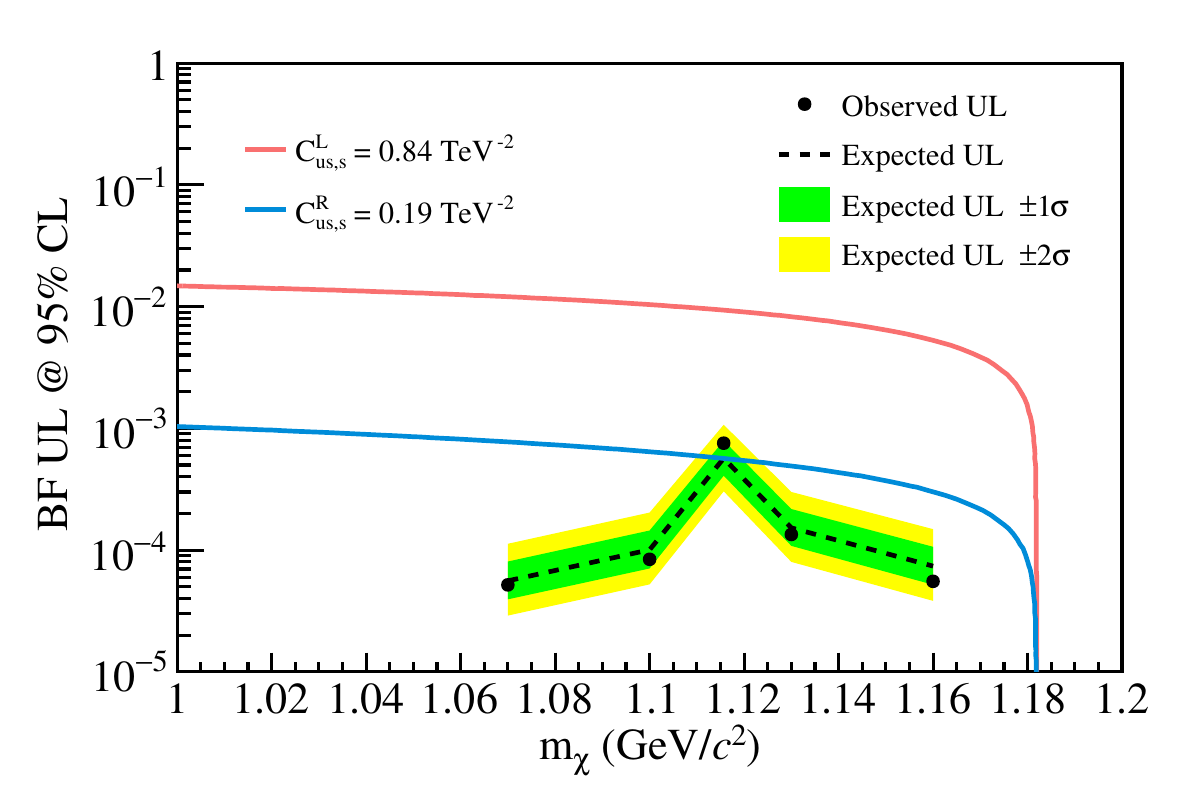}}
        \subfigure[]
        {\includegraphics[width=0.31\textwidth]{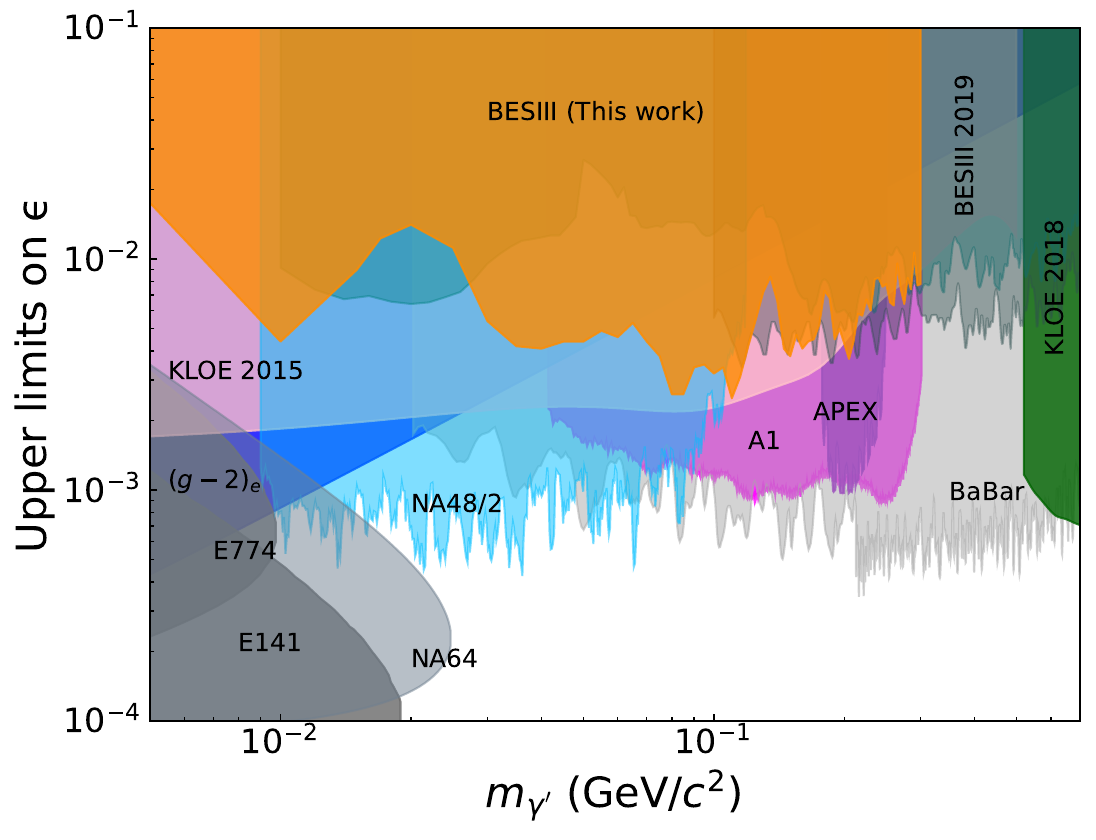}}
        
	\caption{
        (a) The ULs on the BFs of $J/\psi\to\phi X,X\to\rm{invisible}$. (b) The ULs on the BF of $\Xi^-\to\pi^-X,X\to\rm{invisible}$. (b) The UL on the coupling strength between a light vector boson and the SM fermion (in units of $eq_f$).
        } 
	\label{fig:other}
\end{figure*}

The motivation for searching for the process $J/\psi \to \phi + \text{invisible}$ is similar to that for the earlier process $\eta \to \pi^0 + \text{invisible}$. A total of $(8774.0 \pm 39.4) \times 10^6$ $J/\psi$ events are utilized for this search, where the $\phi$ meson is reconstructed from $K^+K^-$ pairs, and the final invisible state is identified by the recoil of $K^+K^-$. To mitigate complex backgrounds, the invariant mass of the final invisible state is required to be less than twice the mass of the $K^0_L$. No significant signal above the expected background is observed in the investigated region, and the UL on the BF is set at $7.0 \times 10^{-8}$ at the 90\% CL. 
ULs at 90\% CL are also provided as a function of the invisible particle mass, ranging from $4 \times 10^{-9}$ to $4 \times 10^{-8}$, as shown in Figure~\ref{fig:other} (a). Additionally, considering that the final invisible state may arise from $\eta$ invisible decay, the UL on the BF for $\eta \to \text{invisible}$ is determined to be $2.4 \times 10^{-5}$, representing an improvement of more than four times compared to the previous best result~\cite{BESIII:2025cyj}.
Currently, there is still a lack of theoretical connection between the measured BF UL and the physical coupling, as this is more theoretically complex in charm decays. Further specific calculations are anticipated.

\vspace{-0.0cm}
\vspace{-0.0cm}

\subsection{Search for dark baryon in $\Xi^-\to\pi^-+\rm{invisible}$}
Dark baryons are theoretical particles that carry baryon number, and several indications support their possible existence. The energy densities ($\rho$) of DM and baryonic matter in the universe are comparable, with $\rho_{\rm{DM}} \sim 5.4\,\rho_{\rm{baryon}}$, which hints at a potential link between their origins and encourages the consideration of dark sector particles that are charged under baryon gauge symmetry. Moreover, neutron lifetime measurements reveal that the lifetime determined via the beam method is longer than that obtained through the storage method, indicating an unknown branching fraction for the decay $n \to \text{dark baryon}$ of approximately $1\%$.
Additionally, examining the decay processes $B \to \text{baryon} + \text{dark baryon}$ and the associated CP violation, the $B$-Mesogenesis mechanism can elucidate the asymmetry between visible matter and antimatter, as well as provide insights into the origin and characteristics of DM~\cite{Alonso-Alvarez:2021oaj}. Naturally, dark baryons are capable of interacting with all flavors of SM quarks, not limited to neutron or $B$ meson decays, thereby offering a valuable avenue for the search for dark baryon particles in hyperon decays at BESIII.

The search for dark baryon particles in the decay $\Xi^- \to \pi^- + \text{invisible}$ is conducted at BESIII for the first time~\cite{BESIII:2025sfl}. This analysis utilizes approximately $10^7$ $\Xi^-\bar{\Xi}^+$ pairs generated from $(10\,084 \pm 44) \times 10^6~J/\psi$ decays. The identification of the $\Xi^-$ candidate involves tagging a $\bar{\Xi}^+$ that decays into $\pi^+\bar{\Lambda}$, with the subsequent decay of $\bar{\Lambda} \to \bar{p}\pi^+$.
The analysis considers various hypotheses for the dark baryon mass, specifically $1.07\,\rm{GeV}$, $1.10\,\rm{GeV}$, $m_\Lambda$, $1.13\,\rm{GeV}$, and $1.16\,\rm{GeV}$, where $m_{\Lambda}$ is the known mass of the $\Lambda$ baryon. Constraints on the dark baryon mass are derived by ensuring high-quality fits in a kinematic analysis, which restricts the invariant mass recoil of the visible particles to align with the dark baryon mass.
The presence of an invisible signal is assessed using the EMC sub-detector. In signal events, no additional hits are recorded in EMC, while SM background events typically show extra hits. No significant signals have been observed. The ULs on the BFs of $\Xi^- \to \pi^- \chi$ are shown in Figure~\ref{fig:other} (b), indicating an improvement over constraints from the LHC as recast in Ref.~\cite{Alonso-Alvarez:2021oaj} within the dark baryon model. Additionally, this analysis may provide constraints relevant to the R-parity violation in supersymmetry models (RPV SUSY).

\subsection{Search for light vector boson in $\chi_{cJ}\to J/\psi V,V\to e^+e^-$}
In addition to the invisible search, BESIII has also conducted a new study on the dark sector decaying to SM particles, specifically in the process $\chi_{cJ} \to J/\psi V$, where $V \to e^+ e^-$. Here, $V$ represents a light vector boson with the interaction Lagrangian:
\begin{eqnarray}
\mathcal{L} \supset \Sigma_f e q_f \epsilon_f V^{\mu} \bar{\psi}_f \gamma_{\mu} \psi_f,
\end{eqnarray}
where $eq_f \epsilon_f$ denotes the coupling strength between the vector boson $V$ and the SM fermion $\psi$, and $eq_f$ is the charge of the fermion. This search is based on $(2712.4 \pm 14.3) \times 10^6$ $\psi(2S)$ events, with the $\chi_{cJ}$ sample obtained from the decay $\psi(2S) \to \gamma \chi_{cJ}$. The signal for the light vector boson is extracted from the distribution of $M_{e^+ e^-}$, and no significant signal is observed. 
The ULs on the coupling strength of $V$, with mass ranging from 5 to 300 MeV, are set to be $(2.5 \sim 17.5) \times 10^{-3}$ (in units of $eq_f$), as shown in Figure~\ref{fig:other} (c). In the dark photon model, $V$ shares a universal $\epsilon_f$ for different fermions. While this search does not provide advantages in the context of the dark photon model, it can uniquely probe the coupling with charm quarks in some non-universal coupling models.

\section{New rare decays searches at BESIII}

\subsection{Search for $\psi(2S)\to e\mu$}
Following the observation of neutrino oscillation, charged lepton flavor violation (CLFV) is permitted in the extended SM, although it is highly suppressed due to the small mass of neutrinos. Various new physics scenarios can enhance the BFs of CLFV decays, such as RPV SUSY, the $Z'$ model, or lepton-quark models. 
A new CLFV study at BESIII is investigating the decay $\psi(2S) \to e\mu$ for the first time. This search is based on $(2367.0 \pm 11.1) \times 10^6$ $\psi(2S)$ events, requiring one electron and one muon in the final state. Eight signal candidates are observed in the data, which is consistent with the estimated background yield of $6.2 \pm 1.4$. The UL on the BF for the decay $\psi(2S) \to e\mu$ is determined to be $1.4 \times 10^{-8}$ at the 90\% CL~\cite{BESIII:2025cdt}.

\subsection{Search for baryon number or lepton number violation processes}
Lepton number (L) and baryon number (B) are always conserved within the SM; however, there are compelling reasons to consider scenarios involving lepton number violation (LNV) and baryon number violation (BNV). The nature of neutrinos, specifically whether they are Dirac or Majorana particles, remains an open question. If neutrinos are Majorana particles, it could lead to the production of LNV processes with $\Delta |L| = 2$. Furthermore, the significant asymmetry between baryons and anti-baryons in the universe suggests the potential existence of BNV. 
There are several new BNV or LNV searches at BESIII, including $J/\psi\to pe^-+c.c.$~\cite{BESIII:2025jtv}, $J/\psi\to K^+K^+e^-e^-+c.c.$~\cite{BESIII:2025wtj}, $\eta\to \pi^+\pi^+e^-e^-+c.c.$~\cite{BESIII:2025ycw}, and $\omega\to\pi^+\pi^+e^-e^-+c.c.$~\cite{BESIII:2025gsy}, with the ULs on the BFs of $3.1\times10^{-8}$, $2.1\times10^{-9}$, $4.6\times10^{-6}$, and $2.8\times10^{-6}$, respectively, all representing the first searches for these decays.

\subsection{Search for charmonium weak decays}
Charmonium weak decays are permitted within the SM, but their BFs are small due to suppression by strong and electromagnetic decays. These decays can be hadronic or semi-leptonic, and the exclusive BFs of the Cabibbo-favored decays are at the order of $10^{-9}$ to $10^{-10}$ in the SM. Certain new physics models, such as the Top-color model or the two-Higgs doublet model, may enhance the BFs of charmonium weak decays.
Recent measurements of charmonium weak decays at BESIII include $J/\psi \to D^-_s e^+ \nu_e + c.c.$~\cite{BESIII:2025frp}, $J/\psi \to D^-_s \rho^+ + c.c.$~\cite{BESIII:2025rjn}, $J/\psi \to D^-_s \pi^+ + c.c.$~\cite{BESIII:2025rjn}, $J/\psi \to D^0 K^{*0} + c.c.$~\cite{BESIII:2025hqb}, $\psi(2S) \to D^-_s \rho^+ + c.c.$~\cite{BESIII:2026lsa}, and $\psi(2S) \to D^-_s \pi^+ + c.c.$~\cite{BESIII:2026lsa}, as summarized in Table~\ref{tab:charmonium_weak_decay}.
\begin{table}[h] 
	\centering
	\caption{The new results of charmonium weak decays from BESIII.}
	\begin{tabular}{*{4}{c}}
		\midrule \midrule
		Channel (+c.c.) & SM BF ($\times10^{-10}$) & New UL on BF & Previous UL on BF\\
		\midrule
		$J/\psi\to D^-_se^+\nu_e$ & $1.9\sim10.21$ & $1.0\times10^{-7}$ & $1.3\times10^{-6}$ \\
        $J/\psi\to D^-_s\rho^+$ & $12.6\sim29.5$ & $8.0\times10^{-7}$ & $1.3\times10^{-5}$ \\
        $J/\psi\to D^-_s\pi^+$ & $2.0\sim7.41$ & $4.1\times10^{-7}$ & $1.3\times10^{-4}$ \\
        $J/\psi\to D^0K^{*0}$ & $1.54\sim7.61$ & $1.9\times10^{-7}$ & $2.5\times10^{-6}$ \\
        $\psi(2S)\to D^-_s\rho^+$ & $12.2$ & $7.0\times10^{-6}$ & $-$ \\
        $\psi(2S)\to D^-_s\pi^+$ & $1.23$ & $1.4\times10^{-6}$ & $-$ \\
		\midrule \midrule
	\end{tabular}
	\label{tab:charmonium_weak_decay}
\end{table}

\section{Summary}
This proceeding presents new results from studies of the dark sector and rare decays at BESIII, yielding stringent constraints on sub-GeV dark matter and establishing first or improved ULs on rare processes involving CLFV, BNV, LNV, and charmonium weak decays. Notably, in the search for $\eta\to\pi^{0}+\mathrm{invisible}$, the dark matter limits derived from BESIII significantly surpass those obtained from direct detection experiments, and also probe the parameter space of the thermal relic DM, underscoring the exceptional potential of BESIII for probing the sub-GeV dark sector. With a large data sample now accumulated in the $\tau$--charm energy region at BESIII, further and more refined results are anticipated in the near future.

\section*{Acknowledgments}

This work is supported in part by National Natural Science Foundation of China (NSFC) under Contracts Nos. 125B2107; National Key R\&D Program of China under Contracts Nos. 2023YFA1606000.

\section*{References}


\end{document}